\documentclass[journal]{IEEEtran}
\usepackage[noadjust]{cite}
\usepackage{graphicx}
\usepackage{float}
\usepackage{color}
\usepackage{amsmath}
\usepackage{enumerate}
\usepackage{amssymb}
\usepackage{fixltx2e}
\usepackage{color}
\usepackage{algorithm}
\usepackage{algorithmic}
\usepackage{multirow}
\usepackage{ragged2e}
\usepackage{makecell}
\usepackage{hhline}

\newcommand{\figref}[1]{Fig.~\ref{#1}}
\usepackage{fancyhdr}

\begin{document}

\title{Antenna Placement Design for Interference Exploitation in Pinching-Antenna Systems\\
%Enhancing Security in RSMA Networks with Cooperative Jamming and Relaying
%\vspace{-0.5cm}
%{\footnotesize (Invited Paper)}
}
%\bibliographystyle{IEEEtran}
%\vspace{-0.3cm}
\author{\normalsize Haoran Pang$^\dag$,  Miaowen Wen$^\ddag$, Fei Ji$^\ddag$, Jun Li$^\ast$, and Yiwei Tao$^\ddag$\\
$^\dag$School of Electronics and Information, 
Guangdong Polytechnic Normal University, Guangzhou 510640, China\\
$^\ddag$School of Electronic and Information Engineering,
South China University of Technology, Guangzhou 510640, China\\
$^\ast$School of Electronics and Communication Engineering, 
Guangzhou University, Guangzhou 510006, China\\
Email: eehrpang@gpnu.edu.cn,  \{eemwwen, eefeiji\}@scut.edu.cn, lijun52018@gzhu.edu.cn, eeyiweitao@mail.scut.edu.cn
%\vspace{-0.3cm}
%	\thanks{
%			Haoran Pang, Fei Ji, and Miaowen Wen are with the School of Electronics and Information Engineering, South China University of Technology, Guangzhou 510641, China (e-mail: eehrpang@mail.scut.edu.cn, \{eefeiji, eemwwen\}@scut.edu.cn).
%%%		\par Lexi Xu is with the
%%%		Research Institute, China United Network Communications Corporation, Beijing 100048, China (e-mail: davidlexi@hotmail.com).
%%%		\par This work was supported in part by the National Natural Science Foundation of China under Grant 61871190, and  in part by the Guangdong Basic and Applied Basic Research Foundation under Grant 2021B1515120067. (\emph{Corresponding author: Miaowen Wen}.) 
%	}
}
% The paper headers
%\markboth{IEEE WIRELESS COMMUNICATIONS LETTERS, VOL. X, NO. X, XXX XXXX}{PANG \MakeLowercase{\textit{et al.}}: Resource Allocation for RSMA-Based Coordinated Direct and Relay Transmission}

\maketitle

\begin{abstract}

%The effectiveness of the proposed schemes is verified through the simulation results.

%Simulation results show that the proposed scheme is an effective enhancement of the security performance of C-RSMA networks. The effectiveness of the proposed scheme is verified through the presentation of the simulation results.
%In addition, both the schemes that the near user only acts as a relay or a jammer are added for comparison with the proposed scheme. Our results demonstrate the potential of the proposed hybrid cooperative user relaying and jamming scheme to enhance the security of C-RSMA networks.
Pinching-antenna systems (PASs) have been proposed as a flexible antenna technology to fulfill the stringent requirements of high data rate and large-scale equipment deployment in future wireless networks. The principle of PA involves mapping a signal over dielectric waveguides for transmission. By adjusting the positions of pinching antennas (PAs) over the waveguides, with the aim of gain enhancement for line-of-sight links and the reduction of large-scale path loss. Symbol-level precoding (SLP) is a nonlinear precoding technique, which converts multi-user interference into constructive interference via beamforming design at symbol level.
%, with the aim of reducing the transmit power and bit error probability.
In this paper, we study the combination of SLP and PAS, leveraging the advantages of PAS to further enhance the ability of SLP to convert constructive interference. The transmit power minimization problem is formulated and solved for the multiple waveguides multiple PAs system by jointly beamforming and PAs' positions design under the SLP principle. The alternating optimization (AO) framework is applied to decouple the beamforming vector and the position coefficient of PA. For the given beamforming vectors, a new objective function is formulated with respect to the positions of the PAs.
% to enhance the received signal strength, which result in further reducing the transmit power at the next iteration. 
 With the characteristics of the formulated objective function, the optimization problem for the position coefficients of PAs can be decomposed into multiple independent subproblems, each corresponding to a PA's position coefficient, and a projected gradient descent (PGD)-based method, constrained by the feasible movable region of each PA, is then developed to obtain the suboptimal position coefficients.
 The performance improvements achieved by the combination of PAS and SLP, as well as the effectiveness of the proposed algorithm are verified through the simulation results.

\end{abstract}


\section{Introduction}
The evolution of next-generation communication systems is driven by increasingly stringent demands for spectral efficiency, latency, and reliability. These imperatives have spurred significant innovation at the physical layer, particularly in channel reconfiguration technologies such as reconfigurable intelligent surfaces (RIS), fluid antennas (FA) , and movable antennas (MA), designed to enhance the wireless propagation environment. RIS consists of numerous passive, low-power reflective elements positioned between the transmitters and receivers \cite{IRS_Overview, IRS_huang}. The phase shift of the reflective elements within the RIS can be adjusted to create a controllable transmission link, thereby enhancing the channel gain. 
FA and MA are channel reconfiguration technologies capable of reshaping the phases of scattering paths between the transmitters and receivers by programmatically adjusting their positions at specific spaces, thereby achieving significant diversity gain \cite{FAtut, MAtut}. Although the aforementioned technologies have been verified capable of enhancing spectral efficiency by tuning specific circuit or geometric parameters, they exhibit limitations in specific scenarios. For instance,  the systems deploying RIS incur the impact of multiplicative fading under the line-of-sight (LoS) blockage or free-space path fading, and the FA or MA-aided systems cannot effectively counteract the effects of large-scale fading by adjusting the positions of antennas at the wavelength-scale level. 

Recently, a novel flexible-antenna system, namely pinching-antenna system (PAS),  has emerged as a complementary channel reconfiguration technology to RIS, FA, and MA, and has thereby attracted considerable attention for its capability to encounter large-scale propagation defects  \cite{PA_first, PA_ding}. PAS consists of a dielectric waveguide connected to each radio-frequency (RF) chain, along with multiple small dielectric elements, referred to as pinching antennas (PAs). The PAs can be  adaptively attached to the waveguide at various positions. Based on the PAS architecture, RF signals propagate through dielectric waveguides with low propagation loss and are radiated by the PAs. By dynamically deploying PAs at appropriate positions along the waveguides, LoS links between transmitters and receivers can be established and enhanced, particularly when PAs are placed close to the receivers \cite{PA_ding}. Additionally, the channel phase associated with each PA varies with its position along the waveguide. This property enables the use of a novel beamforming technique, termed pinching beamforming, which enables flexible and cost-effective multiple-input multiple-output (MIMO) implementations \cite{Pinching_beamforming}. Although PAS is an emerging technology still in its early stages of development, numerous studies across diverse communication scenarios have validated its potential to deliver significant performance gains. In \cite{PA_array}, an array gain maximization problem for the PAS was studied, and the existence of an optimal PA number and the optimal inter-antenna spacing were demonstrated based on the upper bound for the array gain. The studies of \cite{PA_modeling} and \cite{PA_multiuser} investigated the optimization problems in downlink multi-user PAS, respectively. Specifically, the authors in \cite{PA_modeling} solved the joint transmit and pinching beamforming optimization problem under both continuous and discrete
pinching antenna activation scenarios via the proposed penalty-based and zero-forcing (ZF)-based methods. In \cite{PA_multiuser}, the user-fairness problem was solved for three transmission structures in multi-waveguide PAS. Furthermore, the authors of \cite{PA_LosBlockage} evaluated the performance improvement achieved by PASs over conventional antenna systems in multi-user scenarios with LoS blockage.

The aforementioned works on pinching beamforming design are predominantly based on the linear precoding principle. The studies on nonlinear precoding designs for PAS remain limited. Symbol level precoding (SLP) is an effective nonlinear precoding technique designed according to the channel state information (CSI) and transmit modulated symbols \cite{Tutorial_SLP, SLP_PSK}. Based on the precoding design at the symbol level, SLP enables the transformation of multi-user interference into constructive components at the intended receivers, which yields significant gains in its error rate (BER) performance and transmit power efficiency.
Inspired by the performance gains of the PAS and SLP in wireless communication systems, we incorporate them to explore their advantages for improving the system performance further.  By equipping the transmitter side with PAs, the exploitation of the multi-user interference can be promoted by creating and reconfiguring the transmission links. To justify this finding, we explore the application of SLP to the multi-waveguide PASs in this paper. The main contributions are summarized as follows. 
A power minimization problem is formulated by jointly optimizing beamforming vectors and position coefficients of PAs via the SLP principle, subject to the minimum SINR requirement for each user. The alternating optimization (AO) framework is applied to decouple the beamforming vector and the position coefficient of PA. For the PA's position optimization, a new objective function is formulated with respect to the position coefficients of PA. With the characteristics of the formulated objective function, the optimization problem for the position coefficients of PAs can be decomposed into multiple independent subproblems. For each subproblem, a PGD-based algorithm, constrained by the feasible movable region of each PA, is developed to obtain the suboptimal solution. 
Simulation results demonstrate substantial performance improvements achieved by combining PAS and SLP, compared with conventional-antenna SLP schemes.

\textit{Notations:}
The superscripts $(\cdot)^T$ and $(\cdot)^H$ denote transpose and conjugate transpose, respectively. The lowercase and uppercase letters in bold denote vectors and matrices, respectively. 
The operations $\left|\cdot \right| $ and $\left\| \cdot \right\| $ denote the  absolute value and Euclidean norm, respectively. The real and imaginary parts of the complex value are denoted as $\mathrm{Re}\lbrace \cdot\rbrace $ and $\mathrm{Im}\lbrace \cdot\rbrace $, respectively. $\mathbb{C}^{a \times b}$ denotes the set of matrices of dimensions $a\times b$ with complex terms. 
A random vector $\mathbf{x}$ conforming to the circularly symmetric complex Gaussian distribution with mean $\mu$ and a covariance matrix $\mathbf{\Sigma}$ is denoted as $\mathbf{x}\sim\mathcal{CN} \left(\mu, \mathbf{\Sigma} \right)$. 

\section{System Model}
\begin{figure}[t]
	\centering
	\includegraphics[width=3.5in]{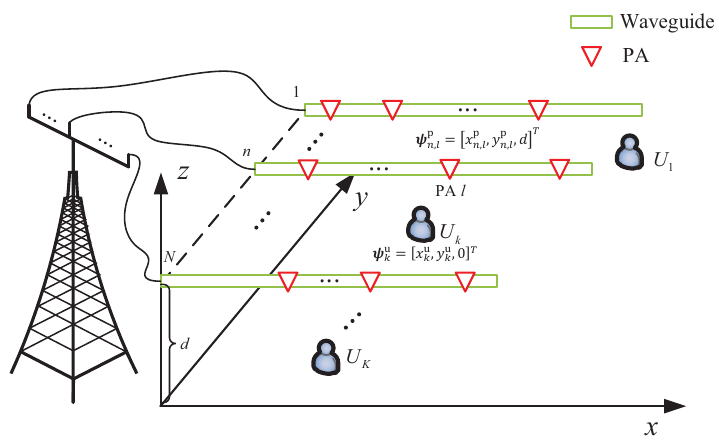}
	\caption{System model of downlink PAS.} \label{systemmodel}
\end{figure}
Consider a PA-enabled downlink multi-user communication system, as shown in \figref{systemmodel},  where a BS equipped with $N$ dielectric waveguides serves $K$ single-antenna users, denoted by $U_{k}$. Each waveguide is connected to a radio frequency (RF) chain and pre-installed $L$ PAs. The indices of waveguides, PAs, and users are collected in $n\in\mathcal{N}=\left\lbrace 1, 2, \ldots, N \right\rbrace $, $l\in\mathcal{L}=\left\lbrace 1, 2, \ldots, L \right\rbrace $, and $k\in\mathcal{K}=\left\lbrace 1, 2, \ldots, K \right\rbrace $, respectively. In \figref{systemmodel}, a three-dimensional Cartesian coordinate system is applied to depict the positions of both waveguides, PAs, and users.  All waveguides are deployed adjacent to each other parallel to the $x$-axis with a height of $d$. Denote the starting point and the position of the $l$-th PA on $n$-th waveguide as $\boldsymbol{\psi}_{n}^{0} = \left[ 0 ,y_{n}^{\mathrm{p}},  d\right]^{T}$ and $\boldsymbol{\psi}_{n,l}^{\mathrm{p}} = \left[ x_{n,l}^{\mathrm{p}}, y_{n}^{\mathrm{p}}, d \right] ^{T}$, respectively. The positions of PAs on the same waveguide are placed in order in this paper, i.e., $x_{n,l+1}^{\mathrm{p}} > x_{n,l}^{\mathrm{p}}$. 
Since all PAs share a single signal propagation path on the same waveguide, a correlated phase-shifted relationship holds between the radiated signals transmitted by different PAs \cite{pozar2012microwave}. Therefore, the phase shift of the signal experienced from the starting point to the PAs over the $n$-th waveguide is given by
\begin{align} \label{phase shift of the signal}
	\mathbf{f}(\mathbf{x}_{n}^{\rm p}) & = \frac{1}{\sqrt{L}}\left[e^{-j\frac{2\pi\left\|\boldsymbol{\psi}_{n,1}^{\mathrm{p}}-  \boldsymbol{\psi}_{n}^{0} \right\| }{\lambda_{g}}}, \ldots, e^{-j\frac{2\pi\left\|\boldsymbol{\psi}_{n,L}^{\mathrm{p}}-  \boldsymbol{\psi}_{n}^{0} \right\| }{\lambda_{g}}} \right]^{T} \nonumber\\
	& = \frac{1}{\sqrt{L}}\left[e^{-j\frac{2\pi x_{n,1}^{\rm p} }{\lambda_{g}}}, \cdots, e^{-j\frac{2\pi x_{n,L}^{\rm p} }{\lambda_{g}}} \right]^T,
\end{align}
where $\mathbf{x}_{n}^{\rm p} = [x_{n,1}^{\rm p}, x_{n,2}^{\rm p}, \ldots, x_{n,L}^{\rm p}]^{T}$. In \eqref{phase shift of the signal},  $\lambda_{g}$ represents the waveguide wavelength, set as $\frac{\lambda}{n_{\rm eff}}$, where $\lambda$ and $n_{\rm eff}$ denote signal wavelength in free space and effective refractive index of dielectric waveguides, respectively. 
The baseband signal radiated from the PAs to the free space is given by
 \begin{align}
	\mathbf{x} = \mathbf{F}(\mathbf{X})\mathbf{W}\mathbf{s},
 \end{align}
where $\mathbf{W} = [\mathbf{w}_{1}, \mathbf{w}_{2},\ldots, \mathbf{w}_{K}]$ with $\mathbf{w}_{k}\in\mathbb{C}^{N\times 1}$ denoting the beamforming vector intended for $U_{k}$, $\mathbf{s}= [s_{1}, s_{2},\ldots, s_{K}]^{T}$ denotes the transmit symbols of the BS for users, which are drawn from the normalized $\mathcal{M}$-ary phase shift keying (PSK) constellation, and 
 \begin{align}
	\mathbf{F}(\mathbf{X}) = 
	\begin{bmatrix}
		\mathbf{f}(\mathbf{x}_{1}^{\rm p}) & \mathbf{0} & \cdots & \mathbf{0}\\
		\mathbf{0} & \mathbf{f}(\mathbf{x}_{2}^{\rm p}) & \cdots & \mathbf{0}\\
		\vdots & \vdots & \ddots & \vdots \\
		\mathbf{0} & \mathbf{0} & \cdots & \mathbf{f}(\mathbf{x}_{N}^{\rm p})
	\end{bmatrix}.
\end{align}
By denoting the position of $U_{k}$ as $\boldsymbol{\psi}_{k}^{\rm u} = [x_{k}^{\rm u}, y_{k}^{\rm u}, 0]$, the channel vector between $n$-th waveguide and $U_{k}$ is given by
 \begin{align}
	\mathbf{h}_{k}(\mathbf{x}_{n}^{\rm p}) = \left[ \frac{\eta e^{-j\frac{2\pi}{\lambda}\left\|\boldsymbol{\psi}_{k}^{\mathrm{u}} - \boldsymbol{\psi}_{n,1}^{\mathrm{p}} \right\| }}{\left\|\boldsymbol{\psi}_{k}^{\mathrm{u}} - \boldsymbol{\psi}_{n,1}^{\mathrm{p}} \right\|}, \cdots, \frac{\eta e^{-j\frac{2\pi}{\lambda}\left\|\boldsymbol{\psi}_{k}^{\mathrm{u}} - \boldsymbol{\psi}_{n,L}^{\mathrm{p}} \right\| }}{\left\|\boldsymbol{\psi}_{k}^{\mathrm{u}} - \boldsymbol{\psi}_{n,L}^{\mathrm{p}} \right\|}\right]^{H},
\end{align}
where $\eta = \frac{c}{4\pi f_{c}}$ represents the pass loss at the reference
distance of 1 m, $c$ is the speed of light, $f_{c}$ is the carrier frequency, and $\scriptstyle \left\|\boldsymbol{\psi}_{k}^{\mathrm{u}} - \boldsymbol{\psi}_{n,l}^{\mathrm{p}} \right\| = \sqrt{\left(x_{k}^{\mathrm{u}}-x_{n,l}^{\mathrm{p}} \right)^2 + \left( y_{k}^{\mathrm{u}}-y_{n}^{\mathrm{p}} \right)^2 + d^2  }$
denotes the Euclidean distance between the $l$-th PA at the $n$-th waveguide and $U_{k}$. 
%and $\left\|\boldsymbol{\psi}_{k}^{\mathrm{u}} - \boldsymbol{\psi}_{n,l}^{\mathrm{p}} \right\|$ can be expressed as
% \begin{align}
%\left\|\boldsymbol{\psi}_{k}^{\mathrm{u}} - \boldsymbol{\psi}_{n,l}^{\mathrm{p}} \right\| = \sqrt{\left(x_{k}^{\mathrm{u}}-x_{n,l}^{\mathrm{p}} \right)^2 + \left( y_{k}^{\mathrm{u}}-y_{n}^{\mathrm{p}} \right)^2 + d^2  }.
%\end{align}
Let $\mathbf{H}_{k}(\mathbf{X}) = [\mathbf{h}_{k}^{H}(\mathbf{x}_{1}^{\rm p}), \mathbf{h}_{k}^{H}(\mathbf{x}_{2}^{\rm p}), \cdots, \mathbf{h}_{k}^{H}(\mathbf{x}_{N}^{\rm p})]$, the received signal at $U_{k}$ is given by
 \begin{align}
	y_{k} 
%	&= \mathbf{H}_{k}(\mathbf{X})\mathbf{F}(\mathbf{X})\mathbf{W}\mathbf{s} +n_{k},\nonumber\\
	 = \mathbf{H}_{k}(\mathbf{X})\mathbf{F}(\mathbf{X})\mathbf{w}_{k}s_{k} + \sum_{i\in\mathcal{K}, i \neq k}\mathbf{H}_{k}(\mathbf{X})\mathbf{F}(\mathbf{X})\mathbf{w}_{i}s_{i} +n_{k},
\end{align}
where $\mathbf{X} = [\mathbf{x}_{1}^{\rm p}, \mathbf{x}_{2}^{\rm p}, \ldots, \mathbf{x}_{N}^{\rm p}]$  and $n_{k}$ represents the additive white Gaussian noise (AWGN) at $U_{k}$ with a mean of $0$ and a variance of $\sigma^2$. Accordingly,
the signal-to-interference-plus-noise ratio (SINR) for decoding the modulated symbol intended for $U_{k}$ can be expressed as 
\begin{align}
{\rm SINR}_{k} &= \dfrac{\left| \mathbf{H}_{k}(\mathbf{X})\mathbf{F}(\mathbf{X})\mathbf{w}_{k} \right|^{2} }{\sum\limits_{i\in\mathcal{K}, i \neq k}\left| \mathbf{H}_{k}(\mathbf{X})\mathbf{F}(\mathbf{X})\mathbf{w}_{i} \right|^{2}+\sigma_{j}^{2}}.
%&=  \dfrac{\left| \sum\limits_{n=1}^{N}\sum\limits_{l=1}^{L}\frac{e^{-j\left( \frac{2\pi}{\lambda}\left\|\boldsymbol{\psi}_{k}^{\mathrm{u}} - \boldsymbol{\psi}_{n,l}^{\mathrm{p}} \right\|  + \frac{2\pi}{\lambda_{g}}x_{n,l}^{\rm p}\right)  }}{\left\|\boldsymbol{\psi}_{k}^{\mathrm{u}} - \boldsymbol{\psi}_{n,l}^{\mathrm{p}} \right\|} w_{k,n} \right|^{2} }{\sum\limits_{i\in\mathcal{K}, i \neq k}\left| \sum\limits_{n=1}^{N}\sum\limits_{l=1}^{L}\frac{e^{-j\left( \frac{2\pi}{\lambda}\left\|\boldsymbol{\psi}_{k}^{\mathrm{u}} - \boldsymbol{\psi}_{n,l}^{\mathrm{p}} \right\|  + \frac{2\pi}{\lambda_{g}}x_{n,l}^{\rm p}\right)  }}{\left\|\boldsymbol{\psi}_{k}^{\mathrm{u}} - \boldsymbol{\psi}_{n,l}^{\mathrm{p}} \right\|} w_{i,n} \right|^{2}+\frac{L}{\eta^2}\sigma_{j}^{2}},
\end{align}
%where $w_{k,n}$ is the $n$-th element of $\mathbf{w}_{k}$.

\section{Problem Formulation and Solution}
The performance of the combination of the SLP and PAS critically depends on the joint design of the beamforming vector and the PA's position. Since the transmit power of the BS is a key performance measure, we focus on the power minimization problem for the considered system. 

Following the SLP principle, the CSI and the modulated symbol information for users are exploited to convert the multi-user interference into constructive interference (CI), which drives the received signal further away from the detection threshold of the signal constellation. This paper employs an $\mathcal{M}$-PSK constellation for symbol modulation. To leverage the geometric principle of CI, the condition $\theta_{AB}\leq \varphi$ ensures that the resultant signal falls within the constructive region, where $\theta_{AB}$ represents the phase angle of the interfering signal and $\varphi = \pi/\mathcal{M}$ \cite{SLP_PSK}. Consequently, the conventional SINR constraints for $U_{k}$ are reformulated as follows:
	\begin{align}
% \left( {\rm Re}(\lambda_{k})-\frac{\sigma_{k}}{\eta}\sqrt{\gamma_{k}L}\right) \tan \theta_{\rm th} -\left| {\rm Im}(\lambda_{k}) \right|  \geq 0  , \forall k\in \mathcal{K},
 \left( {\rm Re}(\lambda_{k})-\sqrt{\gamma_{k}\sigma_{k}^{2}}\right) \tan \theta_{\rm th} -\left| {\rm Im}(\lambda_{k}) \right|  \geq 0  , \forall k\in \mathcal{K},
\end{align}
where $\gamma_{k}$ denotes the minimum required SINR for $U_{k}$, and $\lambda_{k}$ denotes the effect of multi-user interference, which the absolute value represents the received power level of $U_{k}$. 
Accordingly, the joint beamforming and PA's position optimization problem for power minimization with guaranteed CI can be formulated as follows:
\begin{subequations} \label{original problem PA}
	\begin{align}
		\min\limits_{\mathbf{W}, \mathbf{X}, \boldsymbol{\lambda} }&~ \sum_{k=1}^{K}\left\| \mathbf{w}_{k}  \right\|^2\\
		\mathrm{s.t.} ~
		&\mathbf{H}_{k}(\mathbf{X})\mathbf{F}(\mathbf{X})\mathbf{W}\mathbf{s} = \lambda_{k}s_{k}, \forall k\in \mathcal{K}, \label{lambda constraint}\\
		& \left( {\rm Re}(\lambda_{k})-\sqrt{\gamma_{k}\sigma_{k}^{2}}\right) \tan \theta_{\rm th} -\left| {\rm Im}(\lambda_{k}) \right|  \geq 0  , \forall k\in \mathcal{K}, \label{CI constraint}\\
		& 0 \leq x_{n,l}^{\rm p}\leq L^{\rm PA}, \forall n\in \mathcal{N},  \forall l\in \mathcal{L}, \label{position range constraint}\\
		& x_{n,l+1}^{\rm p} - x_{n,l}^{\rm p} \geq  \Delta x^{\rm p}, \forall n\in \mathcal{N},  \forall l\in \mathcal{L}, \label{position gap constraint}
	\end{align}
\end{subequations}
where  $s_{k}$ denotes the modulated symbol intended for $U_{k}$, $\boldsymbol{\lambda} =\left[\lambda_{1}, \lambda_{2}, \ldots \lambda_{K}  \right] $, and $\mathbf{s} = [s_{1}, s_{2}, \ldots, s_{K}]^{T}$. In problem \eqref{original problem PA}, constraint \eqref{position range constraint} restricts the range of PA's position so that it does not exceed maximum length of the waveguide $L^{\rm PA}$, and constraint \eqref{position gap constraint} ensures the interval between PAs is no less than $\Delta x^{\rm p}$. Since the high coupling between $\mathbf{W}$ and $\mathbf{X}$ in problem \eqref{original problem PA}, an AO framework is appropriately applied to decouple this problem in an intuitive manner. For the beamforming optimization with fixed PA's positions, problem \eqref{original problem PA} degrades into a convex problem, which can be solved by using standard optimization tools \cite{boyd2004convex}. 
 
For given $\mathbf{w}_{k}$, the positions of PAs exhibit a weak correlation with the transmit power at the BS, yet are a critical determinant of received signal quality. It can be observed that the optimal $\mathbf{W}$ in problem \eqref{original problem PA} inherently satisfies the inequality conditions of CI constraint \eqref{CI constraint}. Therefore, the values on the left-hand side of constraint \eqref{CI constraint} are promoted by optimizing the PA's location for given $\mathbf{w}_{k}$, which leads us to construct a new optimization problem regarding the PA's location, as follows:
\begin{subequations} \label{new PA problem}
	\begin{align}
		\min\limits_{\mathbf{X} }&~ \sum_{k=1}^{K}\left( \left| {\rm Im}(\lambda_{k}) \right|- \tan \theta_{\rm th}\left( {\rm Re}(\lambda_{k})-\sqrt{\gamma_{k}\sigma_{k}^2}\right)\right) \\
		\mathrm{s.t.} ~ 
		& \eqref{position range constraint}, \eqref{position gap constraint}.
	\end{align}
\end{subequations}
Recall constraint \eqref{lambda constraint}, $\lambda_{k}$ can be rewritten as
	\begin{align}
		\lambda_{k} &= \mathbf{H}_{k}(\mathbf{X})\mathbf{F}(\mathbf{X})\mathbf{W}\mathbf{s} /s_{k}\nonumber\\
		&= \frac{\eta}{s_{k}\sqrt{L}}\sum\limits_{m=1}^{K}\sum\limits_{n=1}^{N}\sum\limits_{l=1}^{L}\frac{e^{-j\left( \frac{2\pi}{\lambda}\left\|\boldsymbol{\psi}_{k}^{\mathrm{u}} - \boldsymbol{\psi}_{n,l}^{\mathrm{p}} \right\|  + \frac{2\pi}{\lambda_{g}}x_{n,l}^{\rm p}\right)  }}{\left\|\boldsymbol{\psi}_{k}^{\mathrm{u}} - \boldsymbol{\psi}_{n,l}^{\mathrm{p}} \right\|} w_{m,n}s_{m},
\end{align}
where $w_{m,n}$ is the $n$-th element of $\mathbf{w}_{m}$.
The amplitude and phase of  $w_{m,n}$ are denoted as $\left| w_{m,n} \right| $ and $\angle w_{m, n}$, respectively. 
%Similarly, the amplitude and phase of $s_{k}$ are 1 and $\angle s_{k}$, respectively.
 Then, we define $q_{n, l, k}(x_{n,l}^{\rm p}) = \scriptstyle\sqrt{\left(x_{k}^{\mathrm{u}}-x_{n,l}^{\mathrm{p}} \right)^2 + \left( y_{k}^{\mathrm{u}}-y_{n,l}^{\mathrm{p}} \right)^2 + d^2  }$ and $f_{n, l,k}(x_{n,l}^{\rm p})\triangleq - \beta_{0}q_{n, l, k}(x_{n,l}^{\rm p}) - \beta_{1}x_{n,l}^{\rm p}$, where $\beta_{0} = \frac{2\pi}{\lambda}$ and $\beta_{1} = \frac{2\pi}{\lambda_{g}}$. Substituting $\left| w_{m,n} \right| $, $\angle w_{m, n}$, $q_{n, l, k}(x_{n,l}^{\rm p})$ and $f_{n, l, k}(x_{n,l}^{\rm p})$ into $\lambda_{k}$, 
 $\lambda_{k}$, ${\rm Im}(\lambda_{k})$, and ${\rm Re}(\lambda_{k})$ can be rewritten as 
 \begin{small}
	\begin{align}
	&\lambda_{k} = \frac{\eta}{\sqrt{L}}\sum_{m=1}^{K}\sum_{n=1}^{N}\sum_{l=1}^{L}\frac{\left|w_{m,n} \right| }{q_{n, l, k}(x_{n,l}^{\rm p})}e^{j\left(f_{n, l, k}(x_{n,l}^{\rm p}) + \beta_{m,n, k}^{\rm ang}\right)},
\end{align}
 \end{small}
  \begin{small}
 	\begin{align}
 		&  {\rm Im}(\lambda_{k}) =  \frac{\eta}{\sqrt{L}}\sum_{m=1}^{K}\sum_{n=1}^{N}\sum_{l=1}^{L}\frac{\left|w_{m,n} \right| }{q_{n, l, k}(x_{n,l}^{\rm p})}\sin\left(f_{n, l, k}(x_{n,l}^{\rm p}) + \beta_{m,n, k}^{\rm ang} \right) \nonumber\\
 		&\qquad \quad \triangleq \frac{\eta}{\sqrt{L}}\sum_{m=1}^{K}\sum_{n=1}^{N}\sum_{l=1}^{L}g^{\rm Im}_{m, n, l, k}(x_{n,l}^{\rm p}) \triangleq G^{\rm Im}_{k}(\mathbf{X}),\\
 		&  {\rm Re}(\lambda_{k}) = \frac{\eta}{\sqrt{L}}\sum_{m=1}^{K}\sum_{n=1}^{N}\sum_{l=1}^{L}\frac{\left|w_{m,n} \right| }{q_{n, l, k}(x_{n,l}^{\rm p})}\cos\left(f_{n, l, k}(x_{n,l}^{\rm p}) + \beta_{m,n, k}^{\rm ang} \right) \nonumber\\
 		&\qquad \quad \triangleq \frac{\eta}{\sqrt{L}}\sum_{m=1}^{K}\sum_{n=1}^{N}\sum_{l=1}^{L}g^{\rm Re}_{m, n, l, k}(x_{n,l}^{\rm p}) \triangleq G^{\rm Re}_{k}(\mathbf{X}), 
 	\end{align}
 \end{small}
respectively, where $\beta_{m,n, k}^{\rm ang} = \angle w_{m,n} +\angle s_{m} - \angle s_{k} $, $\angle s_{m}$ and $\angle s_{k}$ are the phases of $s_{m}$ and $s_{k}$, respectively.
 As a result, problem \eqref{new PA problem} can be reformulated as 
 \begin{small}
	\begin{align}\label{transformed PA problem}
		\min\limits_{\mathbf{X} }&~ 
		\sum_{k=1}^{K}\left( \frac{\eta}{\sqrt{L}}\left( \left| G^{\rm Im}_{k}(\mathbf{X})  \right|-  G^{\rm Re}_{k}(\mathbf{X})\tan \theta_{\rm th}\right)  + \sqrt{\gamma_{k}\sigma_{k}^{2}} \tan \theta_{\rm th}\right) \nonumber\\
		\mathrm{s.t.} ~
		& \eqref{position range constraint}, \eqref{position gap constraint}.
	\end{align}
 \end{small}
 
Following the principle of $\left| a +b \right| \leq \left| a  \right| + \left| b \right|$, we have $\left| G^{\rm Im}_{k}(\mathbf{X})  \right| \leq \sum_{m=1}^{K}\sum_{n=1}^{N}\sum_{l=1}^{L}\left| g^{\rm Im}_{m, n, k}(x_{n,l}^{\rm p})\right| $. Replacing $\left| G^{\rm Im}_{k}(\mathbf{X})  \right|$ with its upper bound , problem \eqref{transformed PA problem} can be decomposed into $N\times L$ sub-problems regarding $\left\lbrace x_{n,l}^{\rm p}\right\rbrace $ by ignoring constraint \eqref{position gap constraint}, where each sub-problem can be given by 
\begin{subequations} \label{transformed PA problem gradient}
	\begin{align}
		\min\limits_{x_{n,l}^{\rm p}}&~ 
		\sum_{k=1}^{K}\sum_{m=1}^{K}\left( \left| g^{\rm Im}_{m, n, l, k}(x_{n,l}^{\rm p})  \right| - g^{\rm Re}_{m, n, l, k}(x_{n,l}^{\rm p})\tan \theta_{\rm th}\right) \label{transformed PA objective function}\\
		\mathrm{s.t.} ~
		& 0 \leq x_{n,l}^{\rm p}\leq L^{\rm PA}.
	\end{align}
\end{subequations}
\noindent To address the challenges posed by the absolute terms in objective function \eqref{transformed PA problem gradient}, the principle $|a+b|=\max\lbrace a+b, -a+b \rbrace $ and the log-sum-exp inequality are applied successively to approximate \eqref{transformed PA objective function} into a smooth function \cite{boyd2004convex}, which results in the following approximated function for \eqref{transformed PA objective function}:
\begin{align}
	&\left| g^{\rm Im}_{m, n, l, k}(x_{n,l}^{\rm p})  \right| - g^{\rm Re}_{m, n, l, k}(x_{n,l}^{\rm p})\tan \theta_{\rm th}\nonumber \\
	&\approx  \varepsilon \log\hspace{-0.1cm}\left(\hspace{-0.1cm}\exp\hspace{-0.1cm}\left(\hspace{-0.1cm} \frac{\bar{\phi}_{m, n, l, k}(x_{n,l}^{\rm p})}{\varepsilon}\hspace{-0.1cm}\right)\hspace{-0.1cm} +\hspace{-0.1cm}\exp\hspace{-0.1cm}\left( \frac{\hat{\phi}_{m, n, l, k}(x_{n,l}^{\rm p})}{\varepsilon}\right)\hspace{-0.1cm} \right)\hspace{-0.1cm}\nonumber\\
    & \triangleq	\Phi_{m, n, l, k}(x_{n,l}^{\rm p}),
\end{align}
where $\varepsilon$ is taken as a sufficiently small positive value, $\bar{\phi}_{m, n, l, k}(x_{n,l}^{\rm p}) \triangleq g^{\rm Im}_{m, n, l, k}(x_{n,l}^{\rm p})   - g^{\rm Re}_{m, n, l, k}(x_{n,l}^{\rm p})\tan \theta_{\rm th}$ and $\hat{\phi}_{m, n, l, k}(x_{n,l}^{\rm p})\triangleq  - g^{\rm Im}_{m, n, l, k}(x_{n,l}^{\rm p})   - g^{\rm Re}_{m, n, l, k}(x_{n,l}^{\rm p})\tan \theta_{\rm th}$.

To satisfy the spacing requirements between PAs, an adjustable moveable region rule for the PAs' positions is proposed in this paper. Assuming the $L^{\rm PA}\gg (K-1)\Delta x^{\rm p}$, the initial movable region for each PA can be set to $L_{\rm init} = \frac{L^{\rm PA} - (L-1)\Delta x^{\rm p}}{L}$. Therefore, the movable region for the $l$-th PA of the $n$-th waveguide is given by
 $ \left[(l-1)\cdot(L_{\rm init} +  \Delta x^{\rm p}), lL_{\rm init} +(l-1) \Delta x^{\rm p}  \right] $. Since each 
position of PA can be achieved by solving sub-problem \eqref{transformed PA problem gradient},  the starting point of the next PA's movable region can be adjusted 
 \begin{algorithm}[H]
	\caption{The PGD-based algorithm for solving subproblem \eqref{transformed PA problem gradient 2}}
	\begin{algorithmic}[1]
		\STATE \textbf{Initialization:} Decompose problem \eqref{transformed PA problem} into $N\times L$ sub-problems. For each sub-problem, set the iteration number $q=1$ and initialized position point $(\mathbf{x}^{\rm p}_{n, l})^{(1)}$.
		\STATE \textbf{Boundary Verification:} Calculate the boundary constraint $\mathcal{C}_{n, l}$ for $\mathbf{x}^{\rm p}_{n, l}$ according to the previously optimized PA's position.
		\STATE \textbf{Repeat:}
		\STATE Compute the gradient of $\boldsymbol{\tilde{\Phi}}_{n, l}((x_{n,l}^{\rm p})^{q})$ according to \eqref{gradient of phi}.
		\STATE Update the step size $\mu_{n, l}^{(q)}$ according to the Armijo-Goldstein criterion.
		\STATE Upadate $(x_{n,l}^{\rm p})^{(q+1)}$ by \eqref{xk update} with $\mu_{n, l}^{(q)}$ and $\triangledown_{x_{n,l}^{\rm p}} 	\boldsymbol{\tilde{\Phi}}_{n, l}((x_{n,l}^{\rm p})^{(q)})$.
		\STATE Project $(x_{n,l}^{\rm p})^{(q+1)}$ according to \eqref{projection}.
		\STATE \textbf{Until:} Convergence or $q=Q_{\max}$.
		\STATE Output the optimal $(x_{n,l}^{\rm p})^{\ast}$.
	\end{algorithmic}
	\label{alg1}
\end{algorithm} 
\noindent based on the previous PA's position. For instance, once the optimal position for $x_{n,l-1}^{\rm p}$ is obtained, the starting point of the movable region for the $l$-the PA at $n$-th waveguide can be updated as $x_{n,l-1}^{\rm p, opt}+\Delta x^{\rm p}$, where $x_{n,l-1}^{\rm p, opt}$ is the solution of the sub-problem regarding $x_{n,l-1}^{\rm p}$. As a result, problem \eqref{transformed PA problem gradient} can be reformulated as 
	\begin{align}\label{transformed PA problem gradient 2}
		\min\limits_{x_{n,l}^{\rm p}}&~ 
		\boldsymbol{\tilde{\Phi}}_{n, l}(x_{n,l}^{\rm p})\triangleq\sum_{k=1}^{K}\sum_{m=1}^{K}\Phi_{m, n, l, k}(x_{n,l}^{\rm p})\nonumber\\
		\mathrm{s.t.} ~ &x_{n,l}^{\rm p}\in\mathcal{C}_{n, l},
	\end{align}
where $\mathcal{C}_{n, l} = \left[ \delta\left( x_{n,l-1}^{\rm p, opt}+\Delta x^{\rm p}\right),  lL_{\rm init} +(l-1) \Delta x^{\rm p}\right] $ , $\delta$ is set to $1$ with $l > 1$, otherwise it is set to $0$. 
Owing to the non-convexity of problem \eqref{transformed PA problem gradient 2}, conventional optimization methods are not directly applicable. Motivated by the simplicity of its constraint, a PGD-based method is developed to yield a suboptimal solution \cite{FAPGD1, FAPGD2}. 
%Problem \eqref{xk optimization problem} is non-convex due to its objective function lacks convexity or concavity, rendering it unsuitable for conventional optimization techniques. 
%It is show that the above problem only includes a simple constraint, which inspires us to apply the PGD method to find the sub-optimal solution. Recently, the application of PGD optimization method has been studied \cite{FAPGD1, FAPGD2}.
%In this paper, we tackle problem \eqref{xk optimization problem} by exploiting the PGD-based method to obtain a suboptimal solution.
The solution to problem \eqref{transformed PA problem gradient 2} achieved by the proposed PGD-based method involves the following two basic steps:

1) Calculate the unconstrained feasible point: The first step is to compute the next unconstrained feasible point. At the $q$-th iteration, the update rule is given by
\begin{align} \label{xk update}
	(x_{n,l}^{\rm p})^{(q+1)} =  (x_{n,l}^{\rm p})^{(q)} - \mu_{n, l}^{(q)}\triangledown_{x_{n,l}^{\rm p}} 	\boldsymbol{\tilde{\Phi}}_{n, l}((x_{n,l}^{\rm p})^{(q)}),
\end{align}
where $\mu_{k}^{(q)}$ is a step size selected according to the Armijo-Goldstein rule, and $\triangledown_{x_{n,l}^{\rm p}} 	\boldsymbol{\tilde{\Phi}}_{n, l}((x_{n,l}^{\rm p})^{(q)})$ denotes the Euclidean gradient of $\boldsymbol{\tilde{\Phi}}_{n, l}((x_{n,l}^{\rm p})^{(q)})$  with respect $(x_{n,l}^{\rm p})^{(q)}$, which is given by \eqref{gradient of phi} at the top of next page. In \eqref{gradient of phi},
\newcounter{mytempeqncnt2}
\setcounter{mytempeqncnt2}{\value{equation}}
\setcounter{equation}{18}
\begin{figure*}[!t]
	\centering % 公式居中
	\begin{align} \label{gradient of phi}
		\triangledown_{x_{n,l}^{\rm p}} \boldsymbol{\tilde{\Phi}}_{n, l}(x_{n,l}^{\rm p}) 
		=\sum_{k=1}^{K}\sum_{m=1}^{K} \dfrac{\exp\left(\bar{\phi}_{m, n, l, k}(x_{n,l}^{\rm p})/ \varepsilon \right)\bar{\phi}^{'}_{m, n, l, k}(x_{n,l}^{\rm p}) + \exp\left(\hat{\phi}_{m, n, l, k}(x_{n,l}^{\rm p})/ \varepsilon \right)\hat{\phi}^{'}_{m, n, l, k}(x_{n,l}^{\rm p})}{\exp\left(\bar{\phi}_{m, n, l, k}(x_{n,l}^{\rm p})/ \varepsilon \right) +  \exp\left(\hat{\phi}_{m, n, l, k}(x_{n,l}^{\rm p})/ \varepsilon \right)},
	\end{align}
	\hrulefill % 添加一条水平线
\end{figure*}
\begin{footnotesize}
\begin{align} \label{derivation of bar}
	&\hspace{-0.1cm}\bar{\phi}^{'}_{m, n, l, k}(x_{n,l}^{\rm p}) \hspace{-0.1cm}= \hspace{-0.1cm}g^{\rm Re}_{m, n, l, k}(x_{n,l}^{\rm p})\hspace{-0.1cm}\left(\hspace{-0.1cm}\frac{(x_{n,l}^{\mathrm{p}}\hspace{-0.1cm}-\hspace{-0.1cm} x_{k}^{\mathrm{u}})\hspace{-0.1cm}\cdot\hspace{-0.1cm}\left(-\beta_{0}q_{n, l, k}(x_{n,l}^{\rm p})\hspace{-0.1cm}+\hspace{-0.1cm}\tan\hspace{-0.05cm} \theta_{\rm th}\right) }{q_{n, l, k}^{2}(x_{n,l}^{\rm p})}\hspace{-0.1cm}-\hspace{-0.1cm} \beta_{1} \hspace{-0.1cm}\right)\hspace{-0.1cm}\nonumber\\
			&-\hspace{-0.1cm} g^{\rm Im}_{m, n, l, k}(x_{n,l}^{\rm p})\hspace{-0.1cm}\left(\hspace{-0.1cm}\frac{(x_{n,l}^{\mathrm{p}}\hspace{-0.1cm}-\hspace{-0.1cm} x_{k}^{\mathrm{u}})\hspace{-0.1cm}\cdot\hspace{-0.1cm}\left(\beta_{0}q_{n, l,k}(x_{n,l}^{\rm p})\tan \hspace{-0.05cm}\theta_{\rm th}\hspace{-0.1cm}+\hspace{-0.1cm}1 \right) }{q_{n, l, k}^{2}(x_{n,l}^{\rm p})}\hspace{-0.1cm}+\hspace{-0.1cm} \beta_{1}\tan\hspace{-0.05cm} \theta_{\rm th} \hspace{-0.1cm}\right),
\end{align}	
\end{footnotesize}
\begin{footnotesize}
	\begin{align} \label{derivation of hat}
		&\hspace{-0.1cm}\hat{\phi}^{'}_{m, n, l, k}(x_{n,l}^{\rm p}) 
		\hspace{-0.1cm}= \hspace{-0.1cm}g^{\rm Re}_{m, n, l, k}(x_{n,l}^{\rm p})\hspace{-0.1cm}\left(\hspace{-0.1cm}\frac{(x_{n,l}^{\mathrm{p}}\hspace{-0.1cm}-\hspace{-0.1cm} x_{k}^{\mathrm{u}})\hspace{-0.1cm}\cdot\hspace{-0.1cm}\left(\beta_{0}q_{n,l,k}(x_{n,l}^{\rm p})\hspace{-0.1cm}+\hspace{-0.1cm}\tan\hspace{-0.05cm} \theta_{\rm th}\right) }{q_{n,l,k}^{2}(x_{n,l}^{\rm p})}\hspace{-0.1cm}+\hspace{-0.1cm} \beta_{1} \hspace{-0.1cm}\right)\hspace{-0.1cm} \nonumber\\
		&-\hspace{-0.1cm} g^{\rm Im}_{m, n, k}(x_{n,l}^{\rm p})\hspace{-0.1cm}\left(\hspace{-0.1cm}\frac{(x_{n,l}^{\mathrm{p}}\hspace{-0.1cm}-\hspace{-0.1cm} x_{k}^{\mathrm{u}})\hspace{-0.1cm}\cdot\hspace{-0.1cm}\left(\beta_{0}q_{n,l,k}(x_{n,l}^{\rm p})\tan \hspace{-0.05cm}\theta_{\rm th}\hspace{-0.1cm}+\hspace{-0.1cm}1 \right) }{q_{n,l,k}^{2}(x_{n,l}^{\rm p})}\hspace{-0.1cm}+\hspace{-0.1cm} \beta_{1}\tan\hspace{-0.05cm} \theta_{\rm th} \hspace{-0.1cm}\right).
	\end{align}	
\end{footnotesize}

2) Project the feasible point: Next, we compute the projection of the next feasible point $(x_{n,l}^{\rm p})^{(q+1)}$ onto the updated movable region, which is given by
\begin{align} \label{projection}
	(x_{n,l}^{\rm p})^{(q+1)}= {\rm Proj}_{\mathcal{C}_{n, l}}((x_{n,l}^{\rm p})^{(q+1)}),
\end{align}
%The update rule for $x_[k]$ at the $q+1$-th iteration is given by
%\begin{align}
%&x_{k}^{(q+1)} =  x_{k}^{(q+1)} + \mu_{k}f_{k}(x_{k}),\\
%&x_{k}^{(q+1)} = \mathcal{P}_{\mathcal{C}_{k}}(x_{k}^{(q+1)}),
%\end{align}
where ${\rm Proj}_{\mathcal{C}}(x)$ is a projection operation ensuring that the next feasible point for the PAs' positions at each iteration always falls in the respective feasible regions, and the rule is given by
\begin{align} 
	{\rm Proj}_{\mathcal{C}}(x):= {\rm arg}\min_{\hat{x}\in\mathcal{C}}\left|x - \hat{x}  \right|.
\end{align}

The overall method for achieving the PAs' positions in this paper can be summarized in Algorithm 1.

\section{Simulation Results}
In this section, computer simulations are conducted to evaluate the transmit power reduction achieved byintegrating SLP with PAS and the proposed PGD-based algorithm for PA's position optimization. In the simulation, users are randomly distributed within a square region with side length $L^{\rm reg} = 20$ meters (m). The BS is located at $[0,L^{\rm reg}/2,0]$, and the number of waveguides is set as $N=4$. Each waveguide is uniformly placed on the square region parallel to the $x$-axis, where the starting point of $n$-th waveguide is set as $[0, (n-1)L^{\rm reg}/(N-1), d]$ with $d=5$ m.  Unless otherwise stated, the system parameters are set as follows: $f_c = 28$ GHz, $n_{eff} = 1.4$, $\sigma^2 = -80$ dBm, $K=4$, and the the modulation constellation is QPSK with $\mathcal{M}=4$. To elucidate the superiority of the proposed algorithm applied for the considered PAS, as compared to the benchmark schemes, including a conventional antenna system and the PAS without adjust the PAs' positions. The PAS with the proposed algorithm for the PA's position design and with random positions of PAs  is abbreviated as ``Proposed scheme'' and ``Random Position'', respectively.  The conventional antenna system via SLP (abbreviated as ``Conventional Antenna '') involves setting antennas gap as $\lambda/2$. The PAS with fixed PA's position is abbreviated as ``Fixed Position'', where  each PA is uniformly distributed along the waveguides.

\begin{figure}
	\centering
	\includegraphics[width=0.5\textwidth]{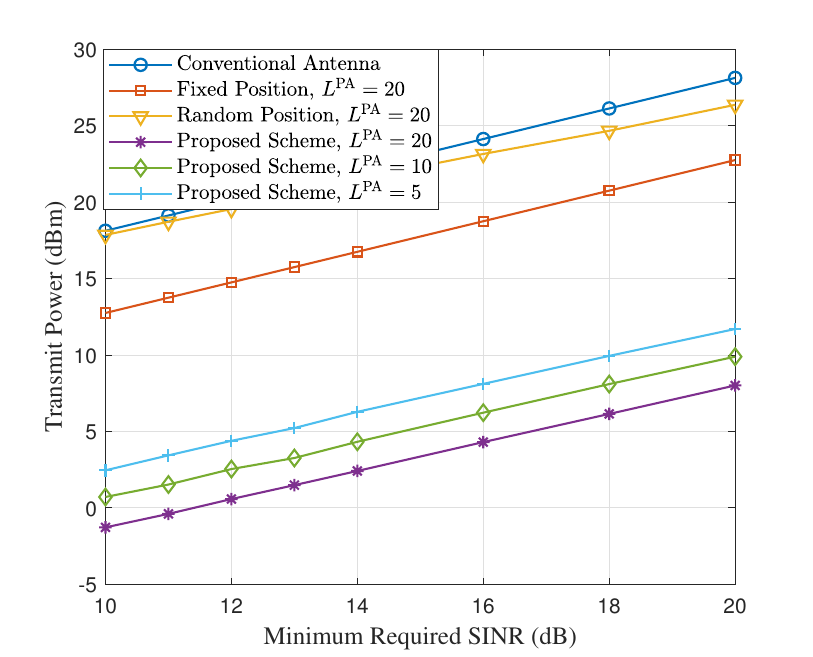}
	\caption{The transmit power versus the minimum required SINR.}	
	\label{powerVsSINR}	
\end{figure}

\figref{powerVsSINR} shows the transmit power versus the minimum required SINR for users, where $L=5$. As expected, the ``Proposed Scheme'' achieves the superior power reduction compared to both the schemes based on the fixed and random positions of the PAs. As the waveguide length increases in the proposed scheme, the required transmit power degrades accordingly.
%In the cases of the proposed scheme with different waveguide lengths, the transmit power degrades with increasing waveguide length. 
Moreover, the ``Fixed Position'' scheme performs better than the ``Conventional Antenna'' scheme. This improvement can be attributed to the phase difference between the PA-emitted signal and the signal from its associated waveguide, which yields an additional pinching beamforming gain. 

\begin{figure}
	\centering
	\includegraphics[width=0.5\textwidth]{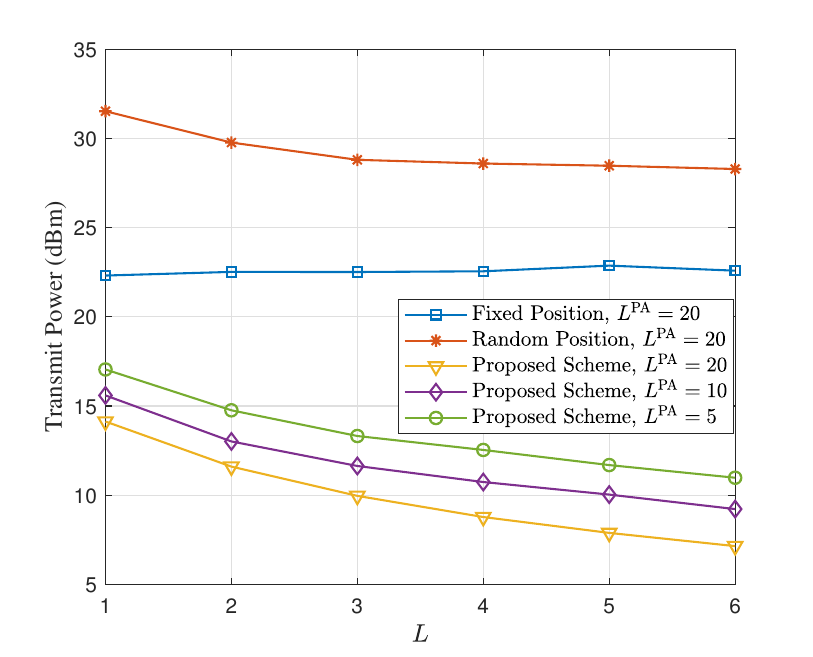}
	\caption{The transmit power versus the number of PAs on each waveguide.}	
	\label{powerVsWDnum}	
\end{figure}

\figref{powerVsWDnum} verifies the transmit power of PAS via SLP with different numbers of PAs on each waveguide, where $\gamma_{k}=20$ dB. As shown in \figref{powerVsWDnum}, the transmit power achieved by the proposed algorithm decreases significantly as the number of PAs at each waveguide increases. 
%due to the improvement of the diversity gain generated by the pinching beamforming. 
The reason is that increasing the number of PAs on each waveguide enhances the diversity gain generated by the pinching beamforming, which results in promoting the transformation of constructive interference. Furthermore, the transmit power achieved by the benchmark schemes with fixed and random PAs' positions varies insignificantly with the number of PAs on each waveguide. Such findings further underscore the importance of developing an efficient algorithm.

\begin{figure}[t]
	\centering
	\includegraphics[width=0.5\textwidth]{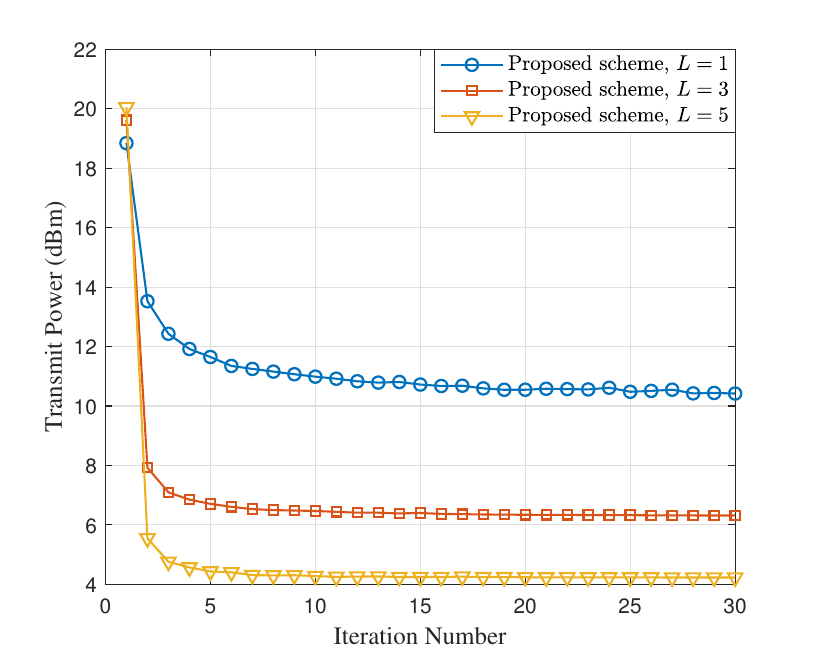}
	\caption{Convergence of the proposed algorithm.}	
	\label{powerVsiteration}	
\end{figure}

\figref{powerVsiteration} exhibits the convergence of the proposed scheme by plotting the transmit power versus the iteration number for the AO framework, where $\gamma_{k}=16$ dB, $L^{\rm PA}=20$ m, and the convergence error is set as $10^{-3}$. The proposed schemes with different numbers of PAs on each waveguide are performed for comparison. It can be observed that the iteration number of the proposed algorithm increases marginally with the rise in the number of PAs on each waveguide, and the algorithm achieves near-optimal performance after approximately 10 iterations for multi-PA cases, which validates its effectiveness.

\section{Conclusion}
In this paper, the downlink multi-user PAS was investigated, where the transmit and pinching beamforming vectors are designed via the SLP principle. We formulated the power minimization problem, and the problem was decoupled into two subproblems by applying AO framework. For the PA's position design, the PGD-based feasible movable region of PAs was developed to achieve the suboptimal solution. Simulation results demonstrated the superiority of the proposed scheme compared with the schemes with fixed position and random position, and the effectiveness of the proposed algorithm.

The proposed algorithm applied in PASs demonstrates superior performance and scalability, with reduced computational complexity and power consumption. However, the PA's position design via the SLP principle requires rapid, continuous adjustments, which necessitates substantial hardware support and incurs high costs, posing a significant challenge for integrating SLP with PAS. The study of long-term and practical SLP solutions for PAS will be an important direction for future research. 

\bibliographystyle{IEEEtran}
\bibliography{mycite}

\vfill

\end{document}